\documentclass[aps,prb,11pt]{revtex4-1}

\usepackage{graphicx}
\usepackage{amsmath}
\usepackage{color}
\begin{document}



\title{Homotopy quantum phase transitions }


\author{Janusz E. Jacak}
\email[]{janusz.jacak@pwr.edu.pl}
\affiliation{Department of Quantum Technologies, Faculty of Fundamental Problems of Technology, Wroc{\l}aw University of Science and Technology, Wyb. Wyspia\'nskiego 27, 50-370 Wroc{\l}aw, Poland}



\date{\today}

\begin{abstract}
We present a new class of quantum phase transitions  that refer neither to local order parameter and critical fluctuations nor to continuous symmetry breaking but are assigned by the step-wise change in topology of the multi-particle system expressed in homotopy terms.
 The energy for  homotopy phases changes  in a step-wise manner and the different homotopies induce specific correlations in planar interacting  multi-particle system representing different commesurability patterns of quantum incompressible states. We illustrate the concept of the homotopy phase transition in the simplest quantum multi-particle system of two repulsing electrons on a 2D finite jellium exposed to a strong perpendicular magnetic field. The homotopy phases related to fractional quantum Hall states  are described and compared with their experimental manifestation.  
\end{abstract}

\pacs{73.22.Pr, 73.43.-f, 05.30.Pr}
\keywords{Path integrals, multiply connected configuration space, multi-particle 2D systems, Bohr-Sommerfeld rule, path homotopy, quantum phase transition}

\maketitle

%
%
%

\section{Introduction}
Phase transitions are convebtionally  classified based on the criterion of  discontinuity of the derivatives of  thermodynamic potentials. The  first-order phase transitions with latent heat are featured by the discontinuity of the first derivative of the free energy, whereas the second-order phase transitions, with discontinuity of the second-order derivative of the free energy. The latter can be accouneted for by the Ginsburg-Landau theory of  phase transitions \cite{phasetr}.
 An arising new  phase at a transition point  is usually assigned by an order parameter (such as density for a gas-liquid transition or magnetization for magnetic transitions) which undergoes a discontinuous step at the first-order phase transition.  At the second-order phase transitions the order parameter continuously  peels off at some critical temperature from its zeroth value in the normal phase. Phase  transitions are typically  associated with a spontaneous break-down of some continuous symmetry. In the ordered  phase with broken continuous symmetry   a Goldstone boson-like collective mode occurs that 
restitutes  the lost symmetry \cite{nambu,goldstone}. Spin waves in an ordered magnetic phase restitute the broken rotational symmetry in ferromagnetic or antiferromagnetic phases \cite{magn}, and phonons restitute broken translational symmetry at liquid-crystal transitions. For superfluid or superconducting phases with broken gauge symmetry \cite{nambu}, the zero sound mode locally restores the broken gauge symmetry (in the case of superfluid He$^3$ with a $3\times 3$ complex order parameter, spin waves also restore the gauge symmetry in the spin channel,  although different than spin waves in magnetic systems \cite{he3,he31}). The second-order phase transitions can be characterized by long-range critical fluctuations of the order parameter, which leads to scaling invariance of the system. For contraction-scaling transformation  a fixed point occurs  according to the Banach theorem. This point is identified  with a transition point and the idea of self-similarity at different scales has been  developed toward the renormalization group approach to second-order phase transitions \cite{wilson}. Due to discontinuity of the order parameter  at the first order transitions with latent heat, the renormalization group approach is not applicable. The melting of crystal is an example of such an transition. This transition is associated with an outcome of income (depending on the direction  of the transition) of some latent heat equivalent to the difference in the thermodynamic potentials of both phases at the transition point. 
 In 2D systems, the second-order phase transitions are excluded by divergence of Goldstone-mode long-range correlations, which destabilizes the new phase \cite{mermin-wagner,hoh}. As an alternative, the idea of a topological transition in 2D was suggested via decoupling  of  vortex and anti-vortex pairs into separated vortices upon the scheme of  Kosterlitz-Touless transitions \cite{kt1,kt2,kt}.

In topologically rich systems other transitions  have also been considered,  neither assigned  by any local order parameter and its fluctuations nor associated with the breaking of any continuous symmetry. 
In the present paper we describe and illustrate   some examples of such transitions in terms of an instant  change of the homotopy classes associated with  particular phases when the external parameters are varied at $T=0$ or even at nonzero  temperature provided that the temperature chaos $kT$ does not exceed  the activation energy of homotopy phases. 

The homotopy related to these phase transitions can be  expressed by the fundamental group, the first homotopy group $\pi_1$, of the configuration space of multiparticle interacting system \cite{mermin1979,rider,spanier1966}. The homotopy group $\pi_1$  collects disjoint classes of trajectories (single-parameter maps) \cite{spanier1966,mermin1979} in this space that cannot be transformed into one another by continuous deformation without cutting  trajectory lines. If a space is assigned with a nontrivial $\pi_1$ group, then such a space is considered as multiply-connected. Otherwise, when $\pi_1=e$ ($e$ is a neutral group element) the space is called as simply-connected. All configuration spaces of multi-particle systems  are multiply-connected, and their nontrivial $\pi_1$ groups are named as braid groups \cite{birman,mermin1979}. Two types of multi-particle systems are considered: with indistinguishable identical particles or distinguishable ones. The former refer to quantum systems, and the corresponding  configuration space of $N$ indistinguishable identical particles has the form $F_N=(M^N-\Delta)/S_N$, where $M^N$ is the $N$-fold normal product of a manifold $M$ where particles are located; $\Delta$ is the  diagonal subset of $M^N$ in which the coordinates of at least two particles coincide and  is subtracted to ensure particle number conservation; and $S_N$ is the permutation group introduced here via a quotient structure to identify points in $F_N$ which differ only by an renumeration of particles. The braid group  $\pi_1(F_N)$ is called the full braid group. The so-called pure braid group defined for classical  distinguishable identical particles  $\pi_1(M^N-\Delta))$  plays the auxiliary role and is a subgroup of the full braid group   \cite{birman,lwitt,mermin1979}.

Multi-particle quantum systems of electrons typical in condensed matter may be also described in  homotopy terms with the utilization of quantum full braid groups. When indistinguishable  particles (electrons) are located in some physical  space, mathematically represented by a manifold $M$, which can be in particular,  an infinite 3D space, 2D (plane), 1D (wire) or a finite sphere, torus or finite 2D plaque, one can in each specific  case identify an appropriate $N$-particle configuration space and its $\pi_1$ group (a  full braid group). The braid groups critically depend on the dimensionality of the manifold $M$ \cite{birman}. For 3D manifolds, the braid groups are always the finite permutation groups $S_N$, whereas for 2D  manifold $M$ (or locally 2D)  the groups are infinite (countable) and much more complicated \cite{birman,jac-ws}. 

\section{Indistinguishability of identical particles in the quantum case}

The prerequisite for any quantum  many-particle system of identical particles is the indistinguishability of particles, which is conventionally referred to  the property resulting from the lack of particle trajectories in the quantum  case \cite{landauqm}. However, the independence of   the indistinguishability of identical particles from conventional  axioms of quantum mechanics is still disputable \cite{landauqm}. This problem is more clearly visible upon the Feynman path formulation of the quantization \cite{feynman1,feynman1964,chaichian1,chaichian2} in which indistinguishability of particles is added in topology-homotopy terms beyond the   quantization scheme for distinguihing particles \cite{wu,lwitt,wilczek,jac-ws}. 
In the case of a multi-particle system of $N$ identical indistinguishing  particles to each
trajectory-bundle of $N$ lines in the $N$-particle configuration space linking an initial  point $\mathbf{r}_1,\dots,\mathbf{r}_N$ on $M$  at a time instant $t$ with a final one $\mathbf{r'}_1,\dots,\mathbf{r'}_N$ at $t'$,  a closed loop of mutually tangled $N$ individual particle trajectory-threads can be attached at any intermediate point. This joined closed trajectory bundle may mix enumeration of particles but not their positions and thus can  represented by an arbitrary element of the full  braid group of the multiparticle system  (as visualized in an example in Fig. \ref{splatanie}).  Because distinct braids are non-homotopic, i.e., are topologically nonequivalent, the domain of the path integral decomposes into a sum of disjoint sectors assigned by the full braid group elements. The discontinuity between these sectors in the path space precludes a definition of a path measure on the entire path space, and this measure must be defined separately in each domain sector. The contributions of all path sectors to the total path integral must be  summed up with  individual unitary factors (due to causality constraints) for  sectors assigned by full  braid group elements. 

 The Feynman path integral  attains thus the form \cite{wu,feynman1964}
\begin{equation}
I(\mathbf{r}_1,\dots, \mathbf{r}_N, t; \mathbf{r}'_1,\dots, \mathbf{r}'_N, t')=\sum_{l\in \pi_1(\Omega)}e^{i\alpha_l}\int d\lambda_l e^{iS[\lambda_l(\mathbf{r}_1,\dots, \mathbf{r}_N, t; \mathbf{r}'_1,\dots, \mathbf{r}'_N, t')]/\hbar},\\
\label{path}
\end{equation}
where $I(\mathbf{r}_1,\dots, \mathbf{r}_N, t; \mathbf{r}'_1,\dots, \mathbf{r}'_N, t')$ is the propagator, i.e., the matrix element of the evolution operator of the total $N$-particle system in the position representation that determines the probability amplitude (complex one) of quantum transition from the multi-particle coordination space  point, $\mathbf{r}_1,\dots, \mathbf{r}_N$, in time instant $t$ to the other point in the configuration space, $\mathbf{r}'_1,\dots, \mathbf{r}'_N$,  in time  instant $t'$. $d\lambda_l$ is the measure in the path space sector enumerated by the $l$-th braid group $\pi_1(F_N)$ element  (braid groups are   countable or finite).  $S[\lambda_l(\mathbf{r}_1,\dots, \mathbf{r}_N, t; \mathbf{r}'_1,\dots, \mathbf{r}'_N, t')]$ is the classical action for the trajectory $\lambda_l$ joining selected  points in the configuration space $F_N$ between time instances $t$, $t'$ and lying in $l$-th sector of the trajectory space with  the $l$th braid loop attached. The whole space of trajectories is decomposed into disjoint sectors enumerated by the braid group element discrete index $l$ (as for a countable group).   It has been proved \cite{lwitt} that the unitary factors (the weights)   associated with contributions of the disjoint sectors of the path integral domain, $e^{i\alpha_l}$ in Eq. ({\ref{path}), establish a one-dimensional unitary representation  (1DUR) of the full  braid group.   Distinct unitary  weights   in the path integral  (i.e., distinct 1DURs of the braid group) determine different types of quantum particles corresponding to the same classical ones. Braids describe particle  exchanges, and thus their  1DURs assign quantum statistics in the system.

Equivalently, the 1DUR  of  a particular braid defines a  phase shift of the multi-particle wave function $\Psi(\mathbf{r}_1,\dots ,\mathbf{r}_N)$ when its arguments $\mathbf{r}_1, \dots, \mathbf{r}_N$ (classical coordinates of particles on the manifold $M$) mutually exchange themselves according to this  braid \cite{sud,imbo} (let us emphasize that  these exchanges are {\it not}  permutations, and the path is important, unless the manifold $M$ is a three- or higher-dimensional space without linear topological defects, such as strings \cite{birman,sud,imbo}).

\begin{figure}[ht]
\centering
\makebox[\linewidth]{
\scalebox{1.0}{\includegraphics{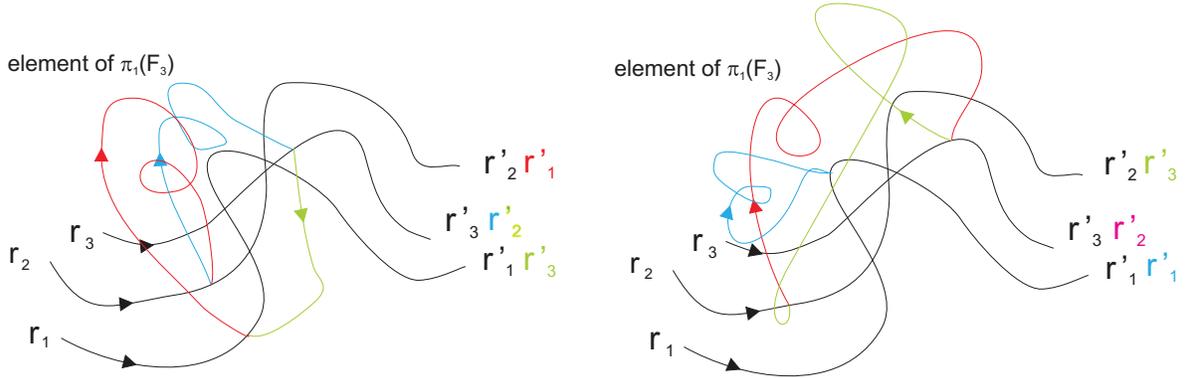}}
}
\caption{\label{splatanie} For a multi-particle trajectory in the configuration space (in the illustration for $N=3$, the configuration space of indistinguishable identical particles is $F_3=(M^3-\Delta)/S_3$), one can add an arbitrary element of the full braid group $\pi_1(F_3)$---indicated by colored tangled lines. Due to nonhomotopy of various braids from the full braid group (linking positions of particles at same intermediate time instant that differ by  permutation)  the trajectories with various braids attached are topologically inequivalent, i.e., cannot be transformed one into another by continuous deformations---they are also nonhomotopic.}
\end{figure} 

For a 3D manifold $M$, full braid groups $\pi_1(F_N)$, are always  finite permutation groups $S_N$. Because the permutation group has only two different 1DURs, $\sigma_i\rightarrow \left\{\begin{array}{l} e^{i0},\\
e^{i\pi}, \end{array}\right.$ (where $\sigma_i$, $i=1,\dots, N-1$, are generators of the permutation group, i.e., exchanges of the $i$th particle with the $(i+1)$-th one, at a certain enumeration of particles), then in 3D  only two quantum statistics are possible, the bosonic and fermionic ones, for $+1$ and $-1$ unitary representations of generators, respectively. When the manifold $M$ is two-dimensional, the full braid group is essentially different than the $S_N$ group. For $M$ being a 2D plane ($R^2$), the full braid group is usually assigned as $B_N$ (Artin group \cite{birman}). It is an infinite group with an infinite number of 1DURs, $\sigma_i\rightarrow e^{i\alpha}, \;\alpha\in[0,2\pi)$, associated with anyons \cite{wilczek}. The generators $\sigma_i$ for the infinite full braid group for a 2D manifold are  exchanges of  particles $i$-th with $(i+1)$-th ones (at certain particle enumeration), but they satisfy different conditions compared to 3D manifold. For the plane $R^2$, the generators of $B_N$ comply with the following conditions:   $\sigma_i \sigma_{i+1} \sigma_i= \sigma_{i+1}\sigma_i\sigma_{i+1}$ for $1\leq i\leq N-2$, and  $\sigma_j\sigma_j=\sigma_j\sigma_i$ for $1\leq i,j\leq N-1,\;|i-j|\geq 2$ (whereas for $S_N$, it was $\sigma_i^2=e$).  

Changes of the braid group causes changes of its 1DUR and of  a related symmetry of the multiparticle wave function and   energies averaged over this function. In a specific 2D case, when   charged interacting $N$-particle system is placed on a surface $S$ (in the thermodynamical limit on whole $R^2$ with kept $\frac{S}{N}=const.$), the strong perpendicular magnetic field may significantly change the  braid group of this system. Although the braids display topology of trajectories and not specific dynamics, the static magnetic field perpendicular to the plane can change the homotopy of all trajectories simultaneously conserving a cylindrical symmetry. This exceptional situation is caused by the  fact that charged classical  particles must move along cyclotron orbits that are of finite size in 2D (in contrast to 3D charged particles for which the drift component of cyclotron movement makes orbits arbitrarily large). For a sufficiently strong magnetic field, the cyclotron orbits of 2D particles with the same  kinetical energy (as in Landau levels or in  other flat bands) may be shorter than the particle separation of electrons on the plane rigidly fixed  by their Coulomb repulsion. This precludes braid exchanges of particles and the full braid group cannot be defined. A braid group must, however, be defined for any correlated multi-particle system, otherwise any quantum statistics cannot be assigned, and any correlated state cannot be organized. 

We have proved that multi-loop braids in 2D can match particles separated by a distance too large for single-particle braids \cite{jac-ws,pra}. These multi-loop braids are $\sigma_i^q$ ($q$ is an integer) and are elements of the original full braid group generated by $\sigma_i$. When $\sigma_i$ cannot be defined, the new braid group (the subgroup of the full braid group) can be defined utilizing $\sigma_i^q$  as its generators (for odd integers $q$, when $\sigma^q_i$ still define exchanges of neighbors, $i$th and $(i+1)$-th particles, what is, however,  not true  for $q$ even integers). These subgroups of the full braid group we call as the cyclotron braid subgroups, and they display distinct homotopy classes for the 2D interacting multiparticle charged system upon a sufficiently strong magnetic field presence. 

However, the implementation of a cyclotron braid subgroup is limited only to a discrete series of magnetic field values in the system of repulsing electrons  with fixed planar concentration (with fixed $N$ and $S$, or in the thermodynamic limit, with fixed $\rho=\frac{N}{S}$). Only for the commensurability of a multi-loop cyclotron braid with particle separation is it possible to define a cyclotron braid subgroup and to  arrange a correlated multi-particle state. Including the possibility of fitting each loop from the multi-loop orbit to particle separation and  taking into account also next-nearest neighboring particles, we obtain the hierarchy of discrete magnetic field values at constant $\rho=\frac{N}{S}$, at which the multi-loop cyclotron braids can be commensurate with the electron distribution homogeneous and steady  due to electron repulsion (electrons are located on the positive uniform jellium). The degeneracy of Landau levels (LLs) of 2D particles in the magnetic field is proportional to the field $B$, $N_0=\frac{BSe}{\hbar}$, and  changes of the LL filling factor, $\nu=\frac{N}{N_0}$, correspond to variation of the magnetic field. The discrete values of $B$ at which cyclotron braid commensurability is admissible define the famous FQHE hierarchy with perfect consistence with experimental observations (ca. 150 various fractional fillings in conventional GaAs and in graphene have been experimentally observed with FQHE thus far \cite{pra,nat,nature}).     

Each commensurability pattern defines a specific homotopy class for trajectories in the considered system upon a rigidly  accommodated magnetic field $B$ (or, equivalently, at a corresponding LL filling factor). The 1DUR of the related specific cyclotron braid subgroup determines the symmetry of the corresponding wave function. In particular, in the lowest Landau level (LLL), these wave functions  can be uniquely  identified,  taking into account that for interacting systems of $N$ electrons in the LLL the wave function must be a  holomorphic function. In the case of the simplest commensurability patterns, one can rederive in this way  the Laughlin functions and next  their generalizations at arbitrary, more complicated multi-loop cyclotron braid commensurability instances (corresponding to different homotopy classes).

For each homotopy  class, the corresponding  wave function attains the form restricted by the cyclotron braid-group symmetry, which leads to a specific   mean energy of this state. This energy is different for various homotopy classes, and the transition between them can be treated as the homotopy phase transition undergoing due to  variation of the magnetic field $B$. The homotopy phases do not  have any order parameters but are distinguished by different patterns of the correlations expressed by the structure of a particular cyclotron braid subgroup. 

The instant change in the commensurability pattern with variation of the magnetic field is particularly notable. The change in homotopy occurs instantly with magnetic field shift when the size of the classical cyclotron orbit changes  in the system of charged repulsing particles with a steady fixed classical distribution on the 2D plane due to the interaction. It must be emphasized that the homotopy phase transitions, although characterized by changes in the classical braid group, are essentially quantum transitions at $T=0$ K corresponding to  the wave function reshaping according the symmetry and unitary representation of the particular cyclotron braid subgroups (and the related step-wise change of the mean energy of interaction).

  The Coulomb-repulsed electrons that are located on the positively charged jellium define a classical Wigner web with the lowest energy distribution of particles with fixed inter-particle spacing (at $T=0$ K). When the temperature grows, the termal chaos $\sim kT$ overcomes the interaction and erodes the classical electron network. Hence, with temperature increase   the commensurability might not be  precisely defined, and the homotopy phase may disappear. The stronger the interaction, the higher  the temperature at which this classical network disintegrates. We see that the interaction does not cause the homotopy phase transitions but is the crucial prerequisite for the commensurability pattern changes with variation of the magnetic field strength.  This relationship agrees with the fact that the  homotopy phase is not assigned by any local order parameter, unlike  to the ordinary phase transitions occurring due to  fluctuations of the order parameter. In particular,  neither the  mean field approach, which highlights the role of interaction in ordinary  phase transitions, nor  the renormalization group approach apply to the presented homotopy phase transitions. In a gas system, the homotopy phase transitions disappear because the braid commensurability loses its meaning when distances between gas particles can be arbitrary as for noninteracting particles. 

The homotopy phase transition is purely  quantum phase  transition (QPT)---it occurs with varying magnetic field at $T=0$ K, although (similarly to other  QPTs) it persists to nonzero temperatures up to a certain temperature when the thermal fluctuations exceed the activation energy of a particular quantum  phase. The activation energy is the energy gain due to homotopy correlations with respect to the uncorrelated phase and depends on the inter-particle interaction and on the interaction with the jellium for a specific homotopy  pattern o of correlations. At higher temperatures, the particle separation is not precisely defined, and the homotopy phase is no longer defined. Thus, for temperatures $kT$ exceeding the activation energies of competing phases the homotopy transition is washed out by the thermal chaos. The ordinary phase transition is a different phenomenon---it is purely classical in terms of thermal fluctuations despite the organization of competing phases is quantum, like e.g., for superfluid or superconducting phases. Conventional QPTs also are governed by fluctuations,   a similar role play here  quantum fluctuations of some order parameter instead of thermal ones. The homotopy transitions are, however, different and are not associated with fluctuations of any local order parameter. They correspond to step-wise changes in the topology of trajectories; i.e., they are assigned to the homotopy transitions in the whole configuration space simultaneously.   

    \section{Examples of homotopy phases}

Let us first consider the integer quantum Hall effect  (IQHE) from the homotopy point of view. This is the correlated state when the LLL filling factor $\nu=\frac{N}{N_0}=1$. The magnetic field $B_0$ corresponding to $\nu=1$  satisfies thus the equation, $N_0=\frac{B_0Se}{h}=N$. In terms of the  cyclotron orbit size, this equality can be rewritten as follows: 
\begin{equation}
\label{iqhe}\frac{S}{N}=\frac{\hbar}{eB_0},
\end{equation}
 where on the left-hand-side of this equation stands $\frac{S}{N}$ which is the sample area per single particle (the measure of particle separation)) and on the right-hand-side, $\frac{\hbar}{eB_0}$,  is the size of a surface of the cyclotron orbit in the LLL. 
 Eq. (\ref{iqhe}) thus determines the homotopy pattern when cyclotron orbits ideally fit to particle separation, allowing braid exchanges, statistics determination  and correlation organization. 

However, if the magnetic field grows, the commensurability condition (\ref{iqhe}) cannot be longer  satisfied. Apparently for $B>B_0$, the commensurability (\ref{iqhe}) is lost, i.e.,  $\frac{S}{N}>\frac{\hbar}{eB}$. In general, there are two possible ways to restore the commensurability necessary to arrange any other correlated state:
\begin{itemize}
    \item to include correlations with next-nearest neighbors---then for the first generation of next-nearest neighbors we obtain on l.h.s instead of $\frac{S}{N}$ the quantity $\frac{S}{N/2}=\frac{2S}{N}$, which, however, only worsens the above inequality (next-nearest neighbor correlations can be, however, convenient  in higher LLs at lower magnetic field, $B<B_0$, and thus larger cyclotron orbits \cite{jet});
\item to include braids with  additional loops---for one additional loop, the simplest such braid is $\sigma_i^3$---as we have proved \cite{pra} (cf. also  paragraph \ref{10}), the  size of the corresponding 3-loop  cyclotron orbit will be $3\frac{\hbar}{eB}$, which restores the commensurability for the field value $B_{1/3}=3B_0$,
\begin{equation}
\label{1/300}
\frac{S}{N}=\frac{3\hbar}{eB_{1/3}},     
\end{equation}
or more generally $\frac{S}{N}=\frac{q\hbar}{eB_{1/q}}$, $q$--odd integer, $B_{1/q}=qB_0$; for this field, we obtain $\nu=\frac{1}{q}$ (in particular $\nu=\frac{1}{3}$ for $q=3$).
 \end{itemize} 

We see that the condition
(\ref{1/300}) defines another homotopy class---the corresponding commensurability pattern holds for 3-loop cyclotron orbits, i.e., braids with one additional loop (because braid to exchange nearest particles utilizes  half of a cyclotron orbit). As the subsequent loops can only be added to braids one by one, $q$ must be an {\it odd} integer: for $q=1$ (i.e., for the ordinary single-loop cyclotron orbit), the corresponding braid takes $1/2$ of this orbit in order to match a neighboring particle; for $q=3$, the corresponding braid takes $3/2$ of the 3-loop cyclotron orbit to match a neighbor; for $q=5$, the braid takes $5/2$ of 5-loop cyclotron orbit, and so on.  

The energies of distinct homotopy phases differ. One can evaluate these energies  by the Metropolis Monte Carlo method \cite{montecarlo1} via   calculation of $<\Psi(z_1,\dots,z_N)|\hat{H}|\Psi(z_1,\dots,z_N))>$ for the corresponding  trial wave functions. For $\nu=\frac{1}{q}$, the Laughlin wave functions are used \cite{laughlin2}:
\begin{equation}
\label{laugh}
\Psi(z_1,\dots,z_N)=A \prod_{i>j}(z_i-z_j)^qe^{-\sum_i |z_i|^2/4l_B^2},   
\end{equation}
$l_B=\sqrt{\frac{\hbar}{eB}}$ is the magnetic length, $z_i=x_i+iy_i$ is the complex notation of $\mathbf{r}_i=(x_i,y_i)$ position of $i$-th particle on the plane, $A$ is the normalization constant. 
The resulted energies of various homotopy phases at distinct $\nu$ are different. Moreover,
any  shift in the magnetic field value, i.e., a shift in $\nu$,  disrupts the commensurability pattern, causing an instant homotopy quantum phase transition.
 The homotopy phase transition is of the topological type without an order parameter, corresponding  to an instant change of the correlation type in the multi-particle system (with $S$ and $N$ constant) induced by changing the cyclotron braid size via the variation  of the magnetic field $B$.

\section{Magnetic field flux quantum in different homotopy phases---the origin of FQHE}

Here we prove that the magnetic field flux quantum changes its value in various homotophy phases. This fact is the origin of the FQHE and was heuristically  modeled \cite{jain} by  auxiliary field quanta attached to electrons to construct a hypothetical composite fermions. Actually none composite fermions exist but  the magnetic field flux quantum changes for various  multiply connected spaces. 
In order to formally  demonstrate the magnetic field  flux quantum modification induced by a specific homotopy class, let us consider the Bohr-Sommerfeld rule, which links the area of the 1D phase space with the corresponding number of quantum states. The quasiclassical wave function in a 1D well, $U(x)$,  with turning points $a$ and $b$ has the form $\Psi(x)=\frac{c}{\sqrt{p}} sin \frac{1}{\hbar}\int_a^xpdx$ for $\Psi(a)=0$  or $\Psi(x)=\frac{c'}{\sqrt{p}} sin \frac{1}{\hbar}\int_b^xpdx$ for $\Psi(b)=0$, where $p(x)=\sqrt{2m(E-U(x))}$ (for simplicity, assuming vertical infinite  borders of the well). Uniqueness of the wave function requires $2 \int_a^bpdx=  \oint p dx=  S_{xp}= n 2\pi \hbar =n h$, which is  the Bohr-Sommerfeld quantization rule (for non-vertical infinite borders, $S_{xp}=(n+\frac{1}{2})h$). The above has been derived upon the condition that the trajectory is single-loop. For a different homotopy class and for a  multi-loop trajectory one obtains, however, 
 2 $ \int_a^bpdx= \oint pdx= S_{px}= (2k+1) n 2\pi \hbar =n (2k+1) h$ for a trajectory  $(a,b)$ with additional $k$ loops.
 Each loop of all $2k$ loops symmetrically pinned  (by $k$) to  both branches, 'upper' ($+p$) and 'lower' ($-p$), of the closed trajectory between $a$ and $b$ in the integral  $\oint pdx$ adds $2\pi$. 
This is of particular importance  when  the Bohr-Sommerfeld rule is applied to a effective 2D phase-space ($P_y,Y$) of $y,x$ components of the 2D kinematic momentum in the presence of a perpendicular magnetic field. The kinematic momentum components, $P_x=-i\hbar\frac{\partial}{\partial x}$ and $P_y=-i\hbar \frac{\partial}{\partial y} -eBx  $  (at the Landau gauge, $\textbf{A}=(0,Bx,0)$) do not commute.  $[P_y,P_x]_-=-i\hbar eB$ and the pair of operators, $Y=\frac{1}{eB}P_x$ and $P_y$, can be treated as operators of canonically conjugated generalized position $Y$ and momentum $P_y$  because    $[P_y,Y]_-=-i\hbar$. Thus,  the 2D effective  phase space $(Y,P_y)$ is actually the $(P_x,P_y)$ space. This 2D kinematic  momentum space is, on the other hand, the  renormalized by the factor $\frac{1}{(eB)^2}$ and turned in plane by $\pi/2$ the ordinary 2D  space $(x,y)$ due to  the quasiclassical formula for the Lorentz force, $\mathbf{F}=\frac{d\mathbf{P}}{dt}=e\frac{d\mathbf{r}}{dt}\times \mathbf{B}$, which gives $dP_{x(y)}=eB dy(-x)$. In the 2D position space, $(x,y)$ trajectories may belong to  different homotopy classes and may be attributed to non-contractible additional loops (as in charged multi-particle planar systems  at sufficiently strong magnetic field). Hence, in this homotopy-rich 2D case, from  the generalized Bohr-Sommerfeld rule, $S_{YP_y}= n (2k+1) h$, or in $(x,y)$  space, $S_{x,y}=\frac{(2k+1)n h}{eB}$, which defines  the quantum of the magnetic field flux, 
\begin{equation}
\label{flux}
\Phi_k=\Delta S_{xy} B=\frac{(2k+1)h}{e}.
\end{equation}
 Only for $k=0$, i.e., for the homotopy class without additional loops, the flux quantum equals $\Phi_0=\frac{h}{e}$. 

Different magnetic field  flux quanta define different size of multi-loop cyclotron orbits. The  IQHE corresponds to $k=0$ (the homotopy class of single-loop cyclotron orbits) and the cyclotron orbit size for $k=0$ equals to  $\Delta S_{xy}= \frac{h}{eB_0}= \frac{S}{N}=\frac{S}{N_0}$, $\nu=\frac{N}{N_0}=1$, ($N_0=\frac{BSe}{h}$ is the LL degeneracy taken here for $B_0$, $S$ is the sample surface size,  $N$ is the number of electrons, $B_0$ is the magnetic field for $\nu=1$).
The FQHE-main line corresponds to $k=1,2,\dots$ (the homotopy class with $q=(2k+1)$-loop cyclotron orbits or braids with $k$ additional loops); e.g., for $k=1$ (the simplest Laughlin state), the three-loop cyclotron orbit has the size $\Delta S_{xy}= \frac{3h}{eB}$. This orbit for $B=3B_0$ fits to interparticle separation $\frac{S}{N}$---hence, from the commensurability condition, $ \frac{3h}{eB}=\frac{S}{N}$, one obtains, $\nu=\frac{N}{N_0}=\frac{N}{BSe/h}=\frac{1}{3}$. It is thus evident that none composite fermions exist but the flux quantum changes. The Laughlin correlations expressed by exponential $q=2k+1$ in the Jastrow polynomial manifesting itselves by the  phase shift $q\pi$ when particle interchange, is not a result of the Aharanov-Bohm phase generated by fluxes attached to composite fermions, but this phase shift is the scalar unitary representation of the braid generator with additional $k$ loops \cite{jac-ws}. The cyclotron braid group theory fully explains FQHE without any fictitious objects like composite fermions with heuristic assertion of somehow pinned flux quanta to each particle. 

It must be emphasized that the Chern-Simons field theory has supported the mystification of composite fermions using the confusing name of gauge field. Despite of this conventional name the Chern-Simon field causes non-canonical transformation of particles and changes their statistics on demand equally artificially as composite fermions do. Thus, the Chern-Simon field theory does not derive composite fermions but offers only a     
field type representation for the auxiliary  fictitious model. In the next paragraphs it is formally proven to which homotopy phases the  composite fermon picture can be applied and which homotopy phases cannot be illustrated by the composite fermions. In aparticular, so-called enigmatic FQHE states in the LLL of monolayer 2DEG systems do not admit composite fermion model as well as  all FQHE states in bilayer systems with tunneling of electrons between layers, as in bilayer graphene \cite{nature,sr1a,bil,bil1} or in closely adjacent GaAs layers \cite{sk1,sk2}. The FQHE states in higher LLs both in Hall monolayers and bilayers are also not of composite fermion character \cite{bil1,nature,amet,jet}.

The quasiclassical method of Bohr-Sommerfeld quantization applied to many particle systems  is interaction independent, i.e., it holds  for arbitrarily strongly interacting multiparticle systems. The sizes of magnetic flux quanta are also interaction independent for different homotopy classes, although the existence of nonhomotopic trajectories in $(x,y)$ space is conditioned by the Coulomb interaction of 2D charged particles. In a gas system of noninteracting particles, their mutual positions are arbitrary, which dismisses correlations and nontrivial homotopies.  
\label{10}
\section{Two-particle illustration  of homotopy classes in 2D}

To better visualize the homotopy phase transition, let us consider the simplest multi-particle, the two-electron 2D system  ($N=2$) located on the surface $S$ (positive jellium) and exposed to a perpendicular  strong magnetic field. The system of two not-bounded electrons has been frequently considered (but without the jellium) because it permits an analytical solution at a countable  series of magnetic field values reflecting hidden symmetries of the related Schr\"odinger equation \cite{pajac,pajac1}. In addition, two electrons in quantum dots were studied, in hyperbolic  \cite{pajac3} or cylindrical dots \cite{pajac2}.

\begin{figure}[ht]
\centering
\resizebox{0.71\textwidth}{!}{\includegraphics{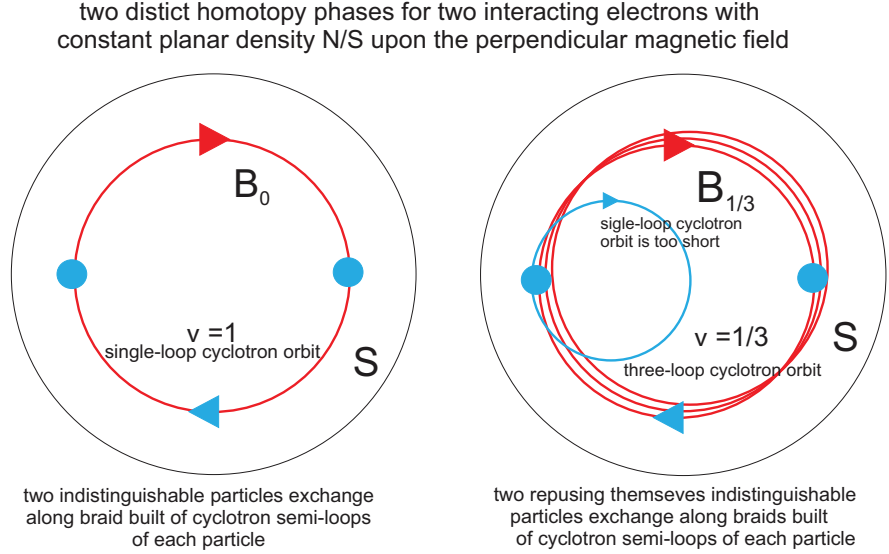}}
\caption{\label{1/3} Example  of two distinct homotopy phases for two repulsing electrons placed on the uniform positive jellium with constant surface density $\frac{N}{S}$ at magnetic field $B_0$ (left) corresponding to $\nu=\frac{N}{N_0}=1$ and $B_{1/3}=3B_0$ (right) corresponding to $\nu=\frac{1}{3}$. The braid trajectory for each electron is a cyclotron semi-loop, with a single-loop cyclotron trajectory for $B_0$ but with a three-loop cyclotron trajectory for $B_{1/3}$ (in the illustration,  the classical cyclotron orbits in the interacting system are schematically presented as circles but may have different shapes with the same surface as circles (e.g., for the commensurability condition $\frac{S}{N}=\frac{\hbar}{eB}$, the single-loop cyclotron orbits with the surface $\frac{\hbar}{eB}$ cannot be circular because it is impossible to fill the plane with circles). For the sake of electrostatic energy minimization, two classical electrons at $T=0$ K are symmetrically positioned on the jellium with radius $r$ at the radius ca. $0.7r$ (the simplest Wigner distribution).}  
\end{figure}

\begin{figure}[ht]
\centering
\resizebox{0.8\textwidth}{!}{\includegraphics{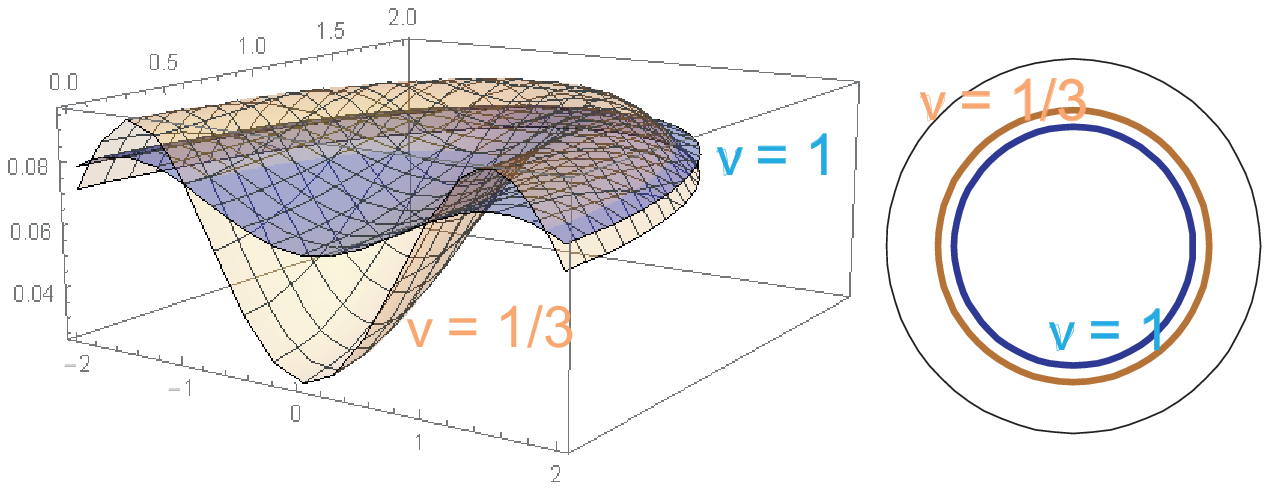}}
\caption{\label{1/33} Single-particle  density of the wave function ($\int d^2r_2 |\Psi_{\nu}(\mathbf{r}_1,\mathbf{r}_2)|^2$, cross-section for better visibility) for $\nu=1$ (blue) and $\nu=\frac{1}{3}$ (brown)---pushing  of charge density out of the center is visible for homotopy class $\nu=\frac{1}{3}$ (left panel). The averaged separation of particles, $\int d^2r_2 \int d^2r_1\Psi_{\nu}^*(\mathbf{r}_1,\mathbf{r}_2)|\mathbf{r}_1-\mathbf{r}_2|
\Psi_{\nu}(\mathbf{r}_1,\mathbf{r}_2))$, is larger for the homotopy class for $\nu=\frac{1}{3}$ (the averaged distance is presented by the circle diameter)  (right panel).} 
\end{figure}

In the case, when the jellium is present,  one can apply the same approach as for many electron 2D systems for IQHE and FQHE.   For the field $B_0$, such that the  the degeneracy of  corresponding LLs, $N_0=\frac{B_0Se}{\hbar}=2$,  two electrons completely fill the LLL, i.e., $\nu=\frac{N}{N_0}=1$. The  related  homotopy phase is defined here by the same  commensurability condition as for IQHE, $\frac{S}{2}=\frac{\hbar}{eB_0}$, which means that the cyclotron braid (half of the cyclotron orbit) perfectly fits the electron separation, as visualized in Fig. \ref{1/3} (left). 
The wave function for this simplest homotopy phase at $B_0$ is $\Psi_1(z_1,z_2)={\cal{A}} (z_1-z_2)e^{(|z_1|^2+|z_2|^2)/4l_{B_0}^2}$, ($z_i=x_i+iy_i$ is a convenient complex representation of the 2D position of the $i$th particle, $\mathbf{r}_i=(x_i,y_i)$, $l_B=\sqrt{\frac{\hbar}{eB}}$ is the magnetic lenght at field $B$). This function is the Laughlin function for $N=2$ and $q=1$, i.e., the Slater function for the completely filled LLL.  It must be emphasized here that this Slater function, being, on the other hand  the wave function of two noninteracting particles in the LLL of gas, is simultaneously the ground state for the interacting  electrons at $\nu=1$. It is an exceptional situation when the different systems, with and without interaction, have the same ground state eigenfunction corresponding, however,  to different Hamiltonians (and different energies), $H_{nint}=\sum_{i=1}^2\frac{(\hat{p}_i-eA_i)^2}{2m}$, $A_i=\frac{1}{2}(-By_i,Bx_i)$ in the symmetrical gauge for $B=B_0$, and   $H_{int}=H_{nint}+\frac{e^2}{4\pi \varepsilon_0 \varepsilon |\mathbf{r}_1-\mathbf{r}_2|} + H_{jj}+H_{ej} $, where the jellium-jellium interaction $H_{jj}= \frac{\rho_0^2}{2} \int_Sd^2r\int_Sd^2r' \frac{e^2}{4\pi \varepsilon_0 \varepsilon |\mathbf{r}-\mathbf{r'}|}$ (with the charge density  $\rho_0=\frac{1}{2\pi l_{B_0}^2}$) and the electron-jellium interaction $H_{el}= -\rho_0 \sum_{i=1}^2 \int_Sd^2r \frac{e^2}{4\pi\varepsilon_0 \varepsilon |\mathbf{r}-\mathbf{r}_i|}$ ($\varepsilon_0$ and $\varepsilon$ are the dielectric constant and the  material permittivity, respectively). The commensurability pattern, $\frac{S}{2}=\frac{\hbar}{eB_0}$, uniquely determines the symmetry of the wave function (in the LLL, the wave function of interacting electrons must be a holomorphic function, defined uniquely by its nodes). Here, the 1DUR of the braid group is $\sigma_i\rightarrow e^{i\alpha}$, where $\alpha= \pi$ is chosen for original fermions, which induces the $z_1-z_2$ polynomial  part. The finite range   exponent $e^{-(|z_i|^2+|z_2|2)/4l_{B_0}^2}$ invariant for particle exchange is  the same common factor in the subspaces of the two-particle Hilbert space spanned by eigenfunctions of noninteracting particles in the LLL for any filling rate (also for $\nu=1$). The resulted two-particle wave function has the form of the Slater function for $N=2$ in the LLL of noninteracting particles. Let us emphasize that this function,  if related to $H_{nint}$,  is not any correlated state (in the gas,  none correlations can be present, and any braid commensurability  also cannot be defined in the gas), whereas the same function if related to $H_{int}$  (as uniquely determined by the braid group 1DUR for $\nu=1$) describes the strongly correlated state of IQHE; this is the simplest homotopy phase. We thus see  that this homotopy correlation is not explicitly built in the wave function form.  Notably, the charge distribution and the averaged electron distance at $\nu=1$ shown in Fig. \ref{1/33} (right) are the same for interacting and noninteracting systems because  the two-particle wave function has the same form in both systems. In the gas case, the finite separation of electrons is caused by fermionic 'repulsion' and is called Pauli virtual crystallization \cite{pc}, although no other homotopic class exists in the gaseous system. In gas any commensurability does not hold and does not impose any restrictions on a full braid group.

The other homotopy phase for two interacting electrons  corresponds e.g.,  to the commensurability pattern $\frac{S}{2}=\frac{3\hbar}{eB_{1/3}}$ at $\nu=\frac{1}{3}$, i.e., to the distinct homotopy class with three-loop cyclotron orbits and thus braids with one additional loop for $B_{1/3}=3B_0$ (cf. Fig. \ref{1/3}, (right)). The Laughlin wave function for this state  state has the form, $\Psi_{1/3}(z_1,z_2)={\cal{B}}(z_1-z_2)^3 e^{-(|z_1|^2+|z_2|^2)/4l_B^2}$.

The energies for both homotopy phases shown above can be calculated directly  as $\Delta E=<\Psi_{\nu}(z_1,z_2)|\hat{H}_{int}-\hat{H}_{nint}|\Psi_{\nu}(z_1,z_2)>$ for $\nu=1$ or $\nu=1/3$. From the Metropolis Monte Carlo estimation we obtain the energies $\frac{\Delta E}{N}=-0.58 \left[\frac{e^2}{4\pi \varepsilon_0 \varepsilon l_B}\right]$ for $\nu=1$, $\frac{\Delta E}{N}=-0.39 \left[\frac{e^2}{4\pi \varepsilon_0 \varepsilon l_B}\right]$ for $\nu=\frac{1}{3}$, and $\frac{\Delta E}{N}=-0.32 \left[\frac{e^2}{4\pi\varepsilon_0 \varepsilon l_B}\right]$ for $\nu=\frac{1}{5}$ (the minus indicates stability, i.e., that  the jellium-electron attraction energy overcomes the jellium-jellium repulsion  and electron-electron repulsion energies). Thus, the energies for distinct homotopy phases are different. 

For both homotopy classes, one can compare the charge density distribution, $\int d^2r_2|\Psi_{\nu}(\mathbf{r}_1,\mathbf{r}_2)|^2$, and the averaged separation of electrons. $\int d^2r_1 \int d^2r_2 \Psi^*_{\nu}(\mathbf{r}_1,\mathbf{r}_2)|\mathbf{r}_1-\mathbf{r}_2|\Psi_{\nu}(\mathbf{r}_1,\mathbf{r}_2)$, for $\nu=1$ and $\nu=\frac{1}{3}$, respectively. As is shown in Fig. \ref{1/33} for multi-loop cyclotron braids  the single-particle charge distribution is pushed out from the  center of the jellium plaque resulting in a larger mean separation of the electron distribution, which  minimizes the electron repulsion energy for the related homotopy class. A comparison of the energies for both phases reveals that the reduction of electron repulsion causes an increase in the electron-jellium interaction, which  prefers a more uniform charge distribution. Both homotopy phases are, however, stable with respect to the nearby quantum state, which cannot be correlated if $\nu$ is slightly shifted out of the homotopy condition.    

These simple examples of the homotopy phases in the case of $N=2$ can be generalized to large electron number systems (also in the thermodynamic limit, provided that the planar density remains constant).

\section{Homotopy phases corresponding to  FQHE hierarchy  in the LLL}

The general form of the braid commensurability in the 2D charged  multi-electron  system in magnetic field
 can be identified via a generalization of the genuine homotopy pattern for IQHE, $\frac{S}{N}=\frac{S}{N_0}$, when the cyclotron orbit size, equal to  $\frac{\Phi_0}{B_0}=\frac{S}{N_0}$ ($N_0=\frac{eB_0S}{h}$ is the LL degeneracy), fits perfectly to the interacting electron separation $\frac{S}{N}$.  At fractional fillings  of the LLL (i.e., at a larger magnetic field $B>B_0$), the cyclotron orbits $\frac{S}{N_0} =\frac{h}{eB}$ are always  smaller than $\frac{S}{N}$ (kept constant here)  and the single-loop cyclotron braids cannot match neighboring electrons. To establish  any correlated state  particle exchanges are, however, necessary to define the statistics of quantum particles via the choice of a braid group 1DUR in the path integral (\ref{path}). Exclusively in 2D, the multi-loop cyclotron orbits have larger sizes compared to the single-loop ones at the same magnetic fields \cite{jac-ws,epl,pra} (cf. the  proof in paragraph \ref{10}). This property is due to the obligatory  distribution of the external field $B$ flux  among all loops of the multi-loop cyclotron orbit in 2D. These loops are all located  in the same plane for the 2D system and  the total flux of the external field passing through the multi-loop orbit is the same as through the single-loop orbit. Hence, the  fraction of this flux falls  per loop (in contrast to a coil in 3D, in which each scroll  adds its own surface, thus enhancing the total flux of an external field passing through a 3D coil). 

The multi-loop braids can thus math neighboring electrons distributed too distantly for braids without additional loops. Therefore, the multi-loop cyclotron orbit commensurability condition attains the following general form, including matching by multi-loop orbits  of nearest or next-nearest electrons:
\begin{equation}
\label{222}
\frac{BS}{N}=(q-1)\frac{h}{ex}\pm\frac{h}{ey},
\end{equation}
where
$q$ is the number of loops  ($q$ must be an odd integer to ensure that the corresponding braid describes particle exchange).  In Eq. (\ref{222}),
$x\geq 1$ (integer) indicates the commensurability of $q-1$ single loops from the $q$-loop cyclotron orbit with every $x$-th particle on the plane ($x=1$ corresponds to nearest neighbors, whereas $x>1$ to next-nearest ones of $x$-th order);
$y\geq x$ (also integer) indicates the commensurability of the last loop of the $q$-loop orbit with every $y$-th particle (next-nearest neighbors for  $y>1$);
$\pm$ indicates the same or opposite (of a figure-eight shape) orientation of the last, i.e., $q$-th, loop.
From (\ref{222}) the following  hierarchy for filling rates is  obtained, 
\begin{equation}
\label{444}
\begin{array}{l}
\nu=\frac{N}{N_0}=\frac{xy}{(q-1)y\pm x},\text{ for LL band electrons},\\
\nu=1-\frac{xy}{(q-1)y\pm x},\text{ for LL band holes}.\\
\end{array} 
\end{equation}
This  general hierarchy reproduces perfectly all experimentally observed filling rates in the LLL for  FQHE \cite{pan2003}. For $x=1$, the hierarchy (\ref{444}) reproduces the conventional CF hierarchy \cite{jain}. For $x > 1$, the  hierarchy (\ref{444})  is beyond the ability  of the CF model  and displays  filling rates for FQHE in the LLL including those  outside the CF hierarchy called as enigmatic states (e.g., $\nu=\frac{5}{13},\frac{4}{11},\frac{3}{8},\frac{3}{10},\frac{5}{17}, \frac{4}{19},\dots$), which are, however,  observed in the experiment in GaAs 2DEG \cite{pan2003}. The comparison of the hierarchy (\ref{444}) with the experimental data is summarized in Fig. \ref{fig1}. 
\begin{figure}[ht]
\centering
\resizebox{1\textwidth}{!}{\includegraphics{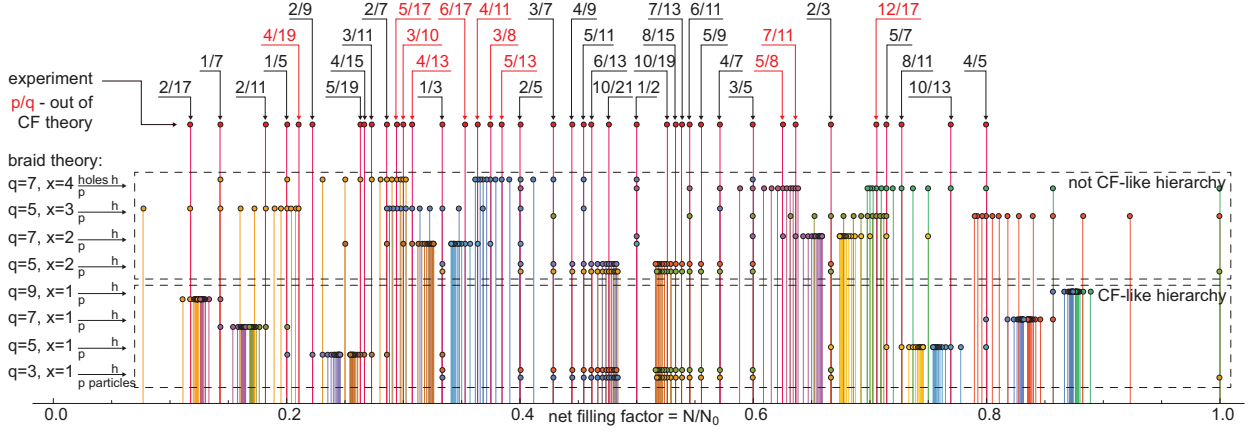}}
\caption{\label{fig1} Comparison of the hierarchy (\ref{444}) with all measured fractional filling rates for FQHE features in the LLL GaAs (spin polarized). The hierarchy series acc. (\ref{444}) for several $y$ are displayed. Filling rates beyond the conventional  hierarchy of  Jain's CFs are shown in red (Hall metal state fraction 1/2 is marked).}
\end{figure}

We note that the CF model agrees with the simplest commensurability case ($x=1$) and breaks down in more complicated commensurability instances as given by Eq. (\ref{444}) for $x>1$. The addition of auxiliary flux quanta to CFs to gain the proper Laughlin phase shift \cite{jain2007} can be interpreted as an effective model of additional loops in braids needed to  reach the nearest neighbors. When the matching of next-nearest electrons is required in some homotopy phase, then such loops (with $x>1$ in Eq. (\ref{444})) are not equivalent to the conventional structure of CFs with pinned flux quanta. In Fig. \ref{fig1}, filling rates beyond the main Jain's CF hierarchy are indicated in red; these filling rates are visible in experiments and all are successfully reproduced by the general hierarchy (\ref{444}) with $x>1$. 
\begin{figure}[ht]
\centering
\resizebox{0.9\textwidth}{!}{\includegraphics{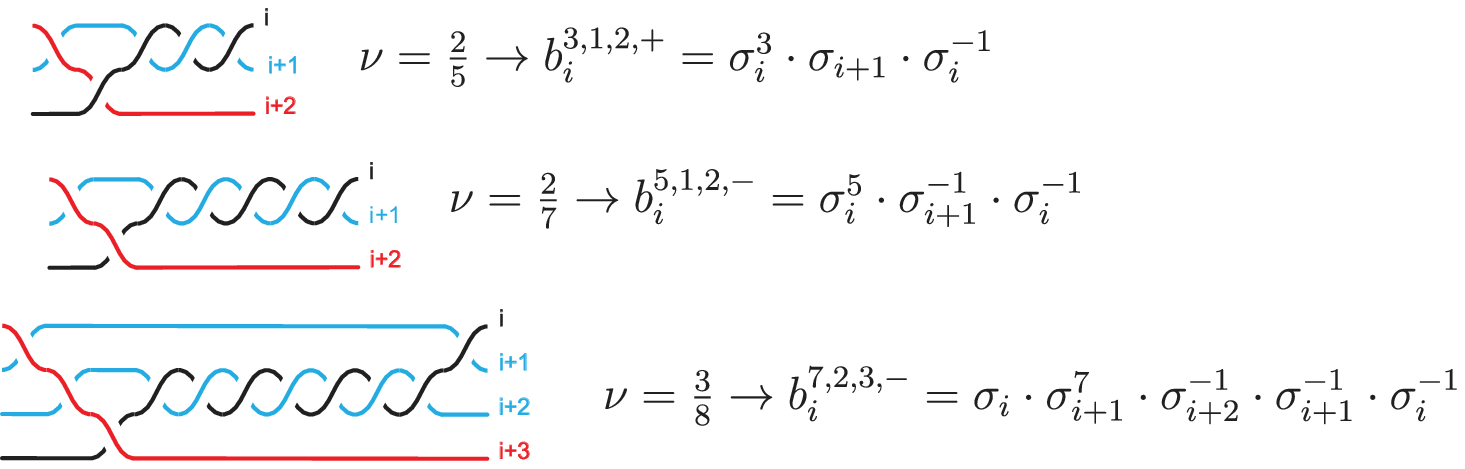}}
\caption{\label{fig3} Examples of the braid cyclotron subgroup generators for several  filling fractions; the third is an example of a filling fraction that cannot be derived using the conventional CF  model (the notation for the generator according to Eq. (\ref{777}) is $b_i^{q,x,y,\pm}$, where $q$ is the number of loops in the cyclotron orbit, $x$ is the next-neighbor order for nesting of the first $(q-1)$ loops, $y$ is the next-neighbor order of the last loop, and $\pm$ corresponds to the same ($+$) or inverted ($-$) orientation of the last loop with respect to the preceding ones).}
\end{figure}

The limit $y\rightarrow \infty$ in Eq. (\ref{444}) displays the general hierarchy of the Hall metal in exactly the same manner as for the archetype of the Hall metal at $\nu=1/2$ (the last orbit, associated with $y$, is then infinite and fits to infinitely distant particles as in the normal Fermi liquid without any magnetic field---this is the Hall metal). The general Hall metal hierarchy in the LLL thus attains the following form:
\begin{equation}
\label{555}
\begin{array}{l}
\nu= \frac{x}{q-1},\text{ for electrons},\\
\nu=1-\frac{x}{q-1},\text{ for holes}. \\
\end{array}
\end{equation}
The Hall metal correlations can manifest themselves at fractions that do not necessarily have even denominators (for $x>1$ even, beyond the conventional Jain's CF  concept of Hall metal, e.g., for $x=2$ in Eq, (\ref{555})). Similarly  the hierarchy (\ref{444}) can display fractions with both odd and even denominators in compliance with the experimental observations \cite{pan2003}. Some fractions are repeated in various lines of the general hierarchy (\ref{444}). This fact  reveals the possibility of realization of  various  commensurability patterns of multi-loop cyclotron orbits with inter-particle spacing $\frac{S}{N}$ at the same $\nu$. The advantage of one commensurability pattern over the others (alternative ones at the same filling rate) is related to energy minimization, i.e., minimization of the Coulomb interaction.

The hierarchy predicted by homotopy phases  reproduces perfectly the experimentally observed series of filling rates for FQHE in the LLL \cite{pan2003} as well as in higher LLs  both in conventional GaAs 2DEG \cite{ll} and in graphene monolayer \cite{nat} and bilayer \cite{nature}. In  bilayer graphene, the homotopy phases successfully explain an unconventional FQHE \cite{bil1,bil,nature,sr1a}, which is out of reach for the CF model because of the specific braid homotopy modification induced by the interlayer hopping of electrons  and cannot be modeled by fictitious flux quanta pinned to electrons \cite{nature}.

\subsection{Comparison of energy for various homotopy phases}
\label{3}

The forms of the cyclotron braid generators for particular homotopy phases permit the construction of trial wave functions that transform themselves according to a selected 1DURs of the cyclotron braid subgroups. In the case of the LLL in which the wave function must be a holomorphic function, the cyclotron braid symmetry determines the wave function unambiguously. For the simplest line of the hierarchy (\ref{444}) with $x=y=1$, i.e., $\nu=\frac{1}{q}, \;q-odd$, the corresponding trial wave functions reproduce the   family of Laughlin functions  \cite{laughlin2} given by Eq. (\ref{laugh}) with the uniquely defined  polynomial parts in the form of the  Jastrow polynomials, $\prod\limits_{i,j,i>j}^{N,N}(z_i-z_j)^q$. The defining characteristic of the Laughlin function is that the $q$-fold zero at each  particle  keeps planar particle density  apart (the stronger the higher $q$ is) and thus diminishes the Coulomb  electron-electron interaction energy, which we have visualized in  the simple  example in Fig. \ref{1/33}. 
The function (\ref{laugh})  transforms in compliance with the 1DUR of the  cyclotron braid subgroup with generators $\sigma_i^q$  i.e., $\sigma_i^q\rightarrow e^{iq\pi}$ (for the fixed 1DUR of the full braid group, $\sigma_i\rightarrow e^{i\alpha}$ with $\alpha =\pi$ choosen to describe  original fermions),  which agrees  with the  Laughlin  phase.

For  more general  homotopy classes corresponding to the hierarchy   (\ref{444}), the generators (elementary braids)  are as follows (for $\pm $ in (\ref{444})):
\begin{equation}
\label{777}
\begin{array}{ll}
b_{i}^{q,x,y, +}=&(\! \sigma_i\! \cdot\! \sigma_{i+\!1}\! \cdot\! ...\! \cdot\! \sigma_{i+x-2}\! \cdot\! \sigma_{i+x-\!1}
\! \cdot\! \sigma_{i+x-2}^{-\!1}\! \cdot\! ...\! \cdot\! \sigma_{i+\!1}^{-\!1}\! \cdot\! \sigma_{i}^{-\!1})^{q-\!1} \\& \cdot\sigma_i\! \cdot\! \sigma_{i+\!1}\! \cdot\! ...\! \cdot\! \sigma_{i+y-2}\! \cdot\! \sigma_{i+y-\!1}\! \cdot\! \sigma_{i+y-2}^{-\!1}\! \cdot\! ...\! \cdot\! \sigma_{i+\!1}^{-\!1}\! \cdot\! \sigma_{i}^{-\!1}\\
\text{and}\\
b_{i}^{q,x,y, -}=&(\! \sigma_i\! \cdot\! \sigma_{i+\!1}\! \cdot\! ...\! \cdot\! \sigma_{i+x-2}\! \cdot\! \sigma_{i+x-\!1}\! \cdot\! \sigma_{i+x-2}^{-\!1}\! \cdot\! ...\! \cdot\! \sigma_{i+\!1}^{-\!1}\! \cdot\! \sigma_{i}^{-\!1})^{q-\!1}\\& \cdot (\! \sigma_i\! \cdot\! \sigma_{i+\!1}\! \cdot\! ...\! \cdot\! \sigma_{i+y-2}\! \cdot\! \sigma_{i+y-\!1}
\! \cdot\! \sigma_{i+y-2}^{-\!1}\! \cdot\! ...\! \cdot\! \sigma_{i+\!1}^{-\!1}\! \cdot\! \sigma_{i}^{-\!1})^{-\!1},\\
\end{array}
\end{equation}
with 1DURs (for $\alpha =\pi$) $ e^{iq\pi}$ (for $+$) and $e^{i(q-2)\pi}$ (for $-)$ (with supplement of the above notation for $x(y)=1$, 
$\sigma_i\cdot\sigma_{i+1}\cdot\dots \cdot\sigma_{i+x-2}\cdot \sigma_{i+x-1}
\cdot \sigma_{i+x-2}^{-1}\cdot \dots \cdot\sigma_{i+1}^{-1}\cdot\sigma_{i}^{-1}=\sigma_i$).
A few  examples of these generators are presented graphically in Fig. \ref{fig3}. These generators induce modifications of the Jastrow polynomials in the following way:
\begin{equation}
\label{888}
\begin{array}{ll}
\Psi^{x,y,+}_q(z_1,z_2,\dots, z_N)=& {\cal{A}}\prod\limits_{\scriptscriptstyle{i,j=1;i<i\;\operatorname{mod}\;x + (j-1)x}}^{N,N/x}(z_i-z_{\scriptscriptstyle{i\;\operatorname{mod}\;x + (j-1)x}})^{q-1}\\ 
& \times \prod\limits_{\scriptscriptstyle{i,j=1;i<i\;\operatorname{mod}\;y + (j-1)y}}^{N,N/y}(z_i-z_{\scriptscriptstyle{i\;\operatorname{mod}\;y + (j-1)y}}) e^{-\sum\limits_i^N\frac{|z_i|^2 }{4l_B^2}},\\ 
\Psi^{x,y,-}_q(z_1,z_2,\dots, z_N)=&
{\cal{A}}\prod\limits_{\scriptscriptstyle{i,j=1;i<i\;\operatorname{mod}\;x + (j-1)x}}^{N,N/x}(z_i-z_{\scriptscriptstyle{i\;\operatorname{mod}\;x + (j-1)x}})^{q-1}\\ 
&\times \prod\limits_{\scriptscriptstyle{i,j=1;i<i\;\operatorname{mod}\;y + (j-1)y}}^{N,N/y}(z_{\scriptscriptstyle{i\;\operatorname{mod}\;y + (j-1)y}} - z_i) e^{-\sum\limits_i^N\frac{|z_i|^2 }{4l_B^2}}.\\
\end{array}
\end{equation} 
From Eq. (\ref{444}) we notice that for $x=1$ it displays the conventional CF  hierarchy. Hence, the above functions correspond to CFs  for $x=1$, as given by Eq. (\ref{999}). It must be emphasized, however, that in the Jain's CF model \cite{jain2007}  trial wave functions are chosen in a different way, utilizing a projection from higher LLs onto the LLL  in order to dispose of poles in higher LL wave function and arrive at a pole-less holomorphic function suitable to the LLL. This procedure is not uniquelly  defined but is rather empirically adjusted to optimize energy, which violates the wave function symmetry in an accidental manner. The  wave functions (\ref{999}) (as a particular case of Eq. (\ref{888})) are, however, not connected with any LL projection and thus are free of the related uncertainty and symmetry perturbations.     
\begin{equation}
\label{999}
\begin{array}{ll}
\Psi^{x=1,y,+}_q(z_1,z_2,\dots, z_N)=&{\cal{}}A\prod\limits_{\scriptscriptstyle{i,j=1,i<j}}^{N,N}(z_i-z_j)^{q-1}\\
&\times \prod\limits_{\scriptscriptstyle{i,j=1;i<i\;\operatorname{mod}\;y + (j-1)y}}^{N,N/y}(z_i-z_{\scriptscriptstyle{i\;\operatorname{mod}\;y + (j-1)y}}) e^{-\sum\limits_i^N\frac{|z_i|^2 }{4l_B^2}},\\
\Psi^{x=1,y,-}_q(z_1,z_2,\dots, z_N)=& {\cal{A}}\prod\limits_{\scriptscriptstyle{i,j=1,i<j}}^{N,N}(z_i-z_j)^{q-1}\\
&\times \prod\limits_{\scriptscriptstyle{i,j=1;i<i\;\operatorname{mod}\;y + (j-1)y}}^{N,N/y}(z_{\scriptscriptstyle{i\;\operatorname{mod}\;y + (j-1)y}} - z_i) e^{-\sum\limits_i^N\frac{|z_i|^2 }{4l_B^2}}.\\
\end{array}
\end{equation} 
The functions (\ref{888}) are proposed as the trial wave functions for correlated states for filling rates (\ref{444}), for which elementary exchanges of particles are defined by braids (\ref{777}) for the case when $x,y>1$. These functions  are similar   to multicomponent Halperin functions \cite{halp}. 

The energy gain in  the homotopy phase  is mostly due to the lowering of the Coulomb repulsion energy $\left<\Psi\right|\sum\limits_{i.j,i>j}^{N,N}\frac{1}{4 \pi \varepsilon_0 \varepsilon }\frac{e^2}{|z_i-z_j|}\left|\Psi\right>$. It is clear that the energy reduction  for function 
(\ref{888}) is weaker at higher $x$ (for the same $q$ and $y$). This  follows from the dilution of correlated particles for larger $x>1$ (the correlation concerns every $x$-th electron only) as expressed in the modified Laughlin-type function (\ref{888}) by reducing the domain of the product. This leads to the diminishing of the repulsion energy gain due to the averaging of the Coulomb energy, $\sum\limits_{i.j,i>j}^{N,N}\frac{e^2}{|z_i-z_j|}$, with the wave function (\ref{888}) instead of (\ref{999}) (or (\ref{laugh})) because $q-1$ fold zero in these functions prevents the approaching of not all electrons in the case of function (\ref{888}) but only its $1/x$ fraction (opposite to the case of function (\ref{laugh}) or (\ref{999}), for which $x=1$). Therefore, states with lower $x$ better reduce electron-electron interaction, and one can expect that states with $x=1$ energetically prevail over states with $ x>1$. To confront the energy values obtained from exact diagonalization for different FQHE fillings, the numerical estimation of energy  for the newly proposed functions (\ref{888}) and (\ref{999}) was performed according to the Monte Carlo  Metropolis scheme \cite{montecarlo1,montecarlo2,metropolis}. Some exemplary results revealing very good overlap with the exact diagonalization  are presented in Table \ref{tab2}.

\begin{table}[th]
\centering
\begin{footnotesize}
\begin{tabular}{p{1.8cm}|p{1.8cm}|p{1.8cm}|p{1.8cm}|p{1.8cm}|p{1.8cm}|p{1.8cm}|p{1.8cm}}
$\nu=N/N_0$&2/5&3/7&4/9&5/11&2/9&3/13&4/17\\
\hline
MMC sim.&$-0.432677$&$-0.441974$&$-0.446474$&$-0.451056$&$-0.342379$&$-0.348134$&$-0.351857$\\
\hline
Ex. diag.&$-0,432804$&$-0.442281$&$-0.447442$&$-0.450797$&$-0.342742$&$-0.348349$&$-0.351189$\\
\end{tabular}
\end{footnotesize}
\caption{Comparison of energy values (per particle in units, $\frac{e^2}{4\pi\varepsilon_0\varepsilon l_B}$)  obtained by exact diagonalization (Ex. Diag.) and by quantum  Monte Carlo simulation (MMC sim.) for some exemplary filling fractions for FQHE (Metropolis Monte Carlo  simulation for the proposed topology-based wave functions, acc. to Eq.  (\ref{999}), for 200 particles).} 
\label{tab2} 
\end{table}

With regard to  multi-loop cyclotron orbits, any particular loop is not  featured (all are equivalent) and thus, in general, each loop can be accommodated to the particle separation independently. Thus, for a $q$-loop orbit, one would deal with the ordered series $x_1\leq x_2\leq \dots \leq x_q$, simplified in (\ref{444}) to $x_1=\dots=x_{q-1}=x, \; x_q=y$. Apparently, the Coulomb repulsion minimization prefers $x_1=\dots=x_{q-1}$, for which the minimization domain is reduced resulting in weaker electron-electron interaction and more efficient energy gain  than for distinct distributions of $x_i$. This explains the choice of the uniform behavior of $q-1$ loops (i.e., $x_1=\dots =x_{q-1}=x$). However,  this is not a rule, and for many fractions, various energetically competitive commensurability opportunities might be considered. 

For a particular filling rate $\nu$, various competing patterns of multi-loop braid commensurability are available in general. Each of the patterns defines a distinct homotopy class with distinct energy (the lowest one defines the ground state at this $\nu$).  Various homotopy patterns at a fixed $\nu$ contribute, however, to the Feynman path integral for a nonstationary case, in which energy is not defined and various trajectory classes leading to different energies must be taken into account upon summation over trajectories in total. We have demonstrated \cite{pra} that he variation of the number of homotopy classes when the magnetic field is shifted reproduces the  longitudinal resistivity $R_{xx}$ experimentally measured in GaAs 2DEG \cite{pan2003}.  The conductivity in the quantum system is proportional to the corresponding path integral, but for a nonstationary case when a distinct number of homotopy classes contributes to the path summation for various $\nu$, the propagator will relatively enhance or diminish in dependence on $\nu $. Resulted variation of the propagator coincides  with local minima ad maxima of $R_{xx}$ experimentally observed \cite{pra}.

\section{Conclusions}

Homotopy topological phases have been identified in 2D charged multi-particle systems exposed to a perpendicular strong magnetic field with multiply connected configuration space. The transition between different homotopy phases does not comply with the conventional phase transition scenario because topological correlations for various braid homotopies are not assigned by any local order parameter. The homotopy class changes instantly according to the cyclotron  braid commensurability pattern for planar interacting multiparticle system, which changes in a step-wise manner with variation of the external magnetic field. Exclusively in 2D multi-particle systems, different multi-loop cyclotron orbits possess  larger sizes compared to the single-loop orbits because of larger magnetic field flux quantum in multiply connected 2D space with homotopy controlled by a magnetic field. This leads to complicated patterns of  multi-loop cyclotron braid commensurability with Coulomb-repulsing 2D electrons homogeneously distributed on uniform positive jellium and including braid nesting with  nearest and next-nearest neighbors. The resulted homotopy class hierarchy agrees with the experimentally observed FQHE hierarchy in conventional GaAs 2DEG and in graphene monolayer and bilayer. The homotopy classes include and with mathematical rigor explain  the heuristic CF approach as the pictorial effective model  of the  simplest instance of braid commensurability concerning nearest neighbors only, which limits the ability of CFs to    
illustrate  FQHE hierarchy. CFs are  usable only  in the LLL of monolayer 2DEG system except of the homotopy classes for so-called enigmatic filling fractions in the LLL when  nesting of braids with next-nearest electrons for first $q-1$ loops of the $q$-loop orbit is required  out of reach for  conventional CF model. CFs fail also in bilayer Hall systems and in higher LLs both of monolayer and bilayer Hall systems. The homotopy classes are characterized by specific quantum statistic  symmetries of  corresponding cyclotron braid group generators defining the polynomial part of the related wave functions and their average energies. These energies agree with the activation energies experimentally observed in Hall experiments and with the energies from exact diagonalization in small models at the corresponding filling rates.   
\begin{acknowledgments}
Supported by the NCN projects P.2011/02/A/ST3/00116 and P.2016/21/D/ST3/00958.

\end{acknowledgments}


\begin{thebibliography}{10}

\bibitem{phasetr}
L.~D. Landau and E.~M. Lifshitz.
\newblock {\em Statistical Physics}.
\newblock Pergamon Press, Oxford, 1969.

\bibitem{nambu}
Y.~Nambu.
\newblock Quasiparticles and gauge invariance in the theory of
  superconductivity.
\newblock {\em Physical Review}, 117:648, 1960.

\bibitem{goldstone}
J.~Goldstone, A.~Salam, and S.~Weinberg.
\newblock Broken symmetries.
\newblock {\em Physical Review}, 127:965, 1962.

\bibitem{magn}
S.~V. Tiablikov.
\newblock {\em Methods in the Quantum Theory of Magnetism}.
\newblock Springer, New York, 1967.

\bibitem{he3}
A.~Leggett.
\newblock A theoretical description of the new phases of liquid {H}e-3.
\newblock {\em Rev. Mod. Phys.}, 48:357, 1976.

\bibitem{he31}
J.~Czerwoko.
\newblock Spin susceptibility of the pseudoisotropic phase of superfluid {H}e-3
  in the acoustic limit (spin waves).
\newblock {\em Journal of Exp. and Theor. Phys.}, 44:575, 1976.

\bibitem{wilson}
K.~G. Wilson and J.~Kogut.
\newblock The renormalization group and the e expansion.
\newblock {\em Physics Reports}, 12:75, 1974.

\bibitem{mermin-wagner}
N.~D. Mermin and H.~Wagner.
\newblock Absence of ferromagnetism or antiferromagnetism in one- or
  two-dimensional isotropic {H}eisenberg models.
\newblock {\em Phys. Rev. Lett.}, 17:1133, 1966.

\bibitem{hoh}
P.~C. Hohenberg.
\newblock Existence of long-range order in one and two dimensions.
\newblock {\em Phys. Rev.}, 158:383, 1967.

\bibitem{kt1}
W.~L. Berezinskii.
\newblock Destruction of long-range order in one-dimensional and
  two-dimensional systems having a continuous symmetry group i, classical
  systems.
\newblock {\em Journal of Exp. Teor. Phys.}, 32, 493 1970.

\bibitem{kt2}
W.~L. Berezinskii.
\newblock Destruction of long-range order in one-dimensional and
  two-dimensional systems possessing a continuous symmetry group. ii. quantum
  systems.
\newblock {\em Journal of Exp. Teor. Phys.}, 61, 1144 1971.

\bibitem{kt}
J.~Kosterlitz and D.~J. Thouless.
\newblock Ordering, metastability and phase transitions in two-dimensional
  systems.
\newblock {\em Journal of Physics C}, 6:1181, 1973.

\bibitem{mermin1979}
N.~Mermin.
\newblock The topological theory of defects in ordered media.
\newblock {\em Rev. Mod. Phys.}, 51:591, 1979.

\bibitem{rider}
L.~H. Ryder.
\newblock {\em Quantum Field Theory, 2nd ed.}
\newblock Cambridge University Press, Cambridge, 1996.

\bibitem{spanier1966}
E.~Spanier.
\newblock {\em Algebraic topology}.
\newblock Springer-Verlag, Berlin, 1966.

\bibitem{birman}
J.~S. Birman.
\newblock {\em Braids, Links and Mapping Class Groups}.
\newblock Princeton UP, Princeton, 1974.

\bibitem{lwitt}
M.~G. Laidlaw and C.~M. DeWitt.
\newblock Feynman functional integrals for systems of indistinguishable
  particles.
\newblock {\em Phys. Rev. D}, 3:1375, 1971.

\bibitem{jac-ws}
J.~Jacak, R.~Gonczarek, L.~Jacak, and I.~J{\'o}{\'z}wiak.
\newblock {\em Application of Braid Groups in {2D} Hall System Physics:
  Composite Fermion Structure}.
\newblock World Scientific, Singapore, 2012.

\bibitem{landauqm}
L.D. Landau and E.~M. Liifsitz.
\newblock {\em Quantum mechanics, no-relativistic theory}.
\newblock Pergamon Press, Oxford, 1965.

\bibitem{feynman1}
R.~P. Feynman.
\newblock {\em The principle of least action in quantum mechanics}.
\newblock Ph.D. thesis, Princeton University, 1942.

\bibitem{feynman1964}
R.~P. Feynman and A.~R. Hibbs.
\newblock {\em Quantum Mechanics and Path Integrals}.
\newblock McGraw-Hill, New York, 1964.

\bibitem{chaichian1}
M.~Chaichian and A.~Demichev.
\newblock {\em Path Integrals in Physics Volume I Stochastic Processes and
  Quantum Mechanics}.
\newblock IOP Publishing Ltd, Bristol; Philadelphia, 2001.

\bibitem{chaichian2}
M.~Chaichian and A.~Demichev.
\newblock {\em Path Integrals in Physics Volume II Quantum Field Theory,
  Statistical Physics and other Modern Applications}.
\newblock IOP Publishing Ltd, Bristol; Philadelphia, 2001.

\bibitem{wu}
Y.~S. Wu.
\newblock General theory for quantum statistics in two dimensions.
\newblock {\em Phys. Rev. Lett.}, 52:2103, 1984.

\bibitem{wilczek}
F.~Wilczek.
\newblock {\em Fractional Statistics and Anyon Superconductivity}.
\newblock World Scientific, Singapore, 1990.

\bibitem{sud}
E.~C.~G. Sudarshan, T.~D. Imbo, and T.~R. Govindarajan.
\newblock Configuration space topology and quantum internal symmetries.
\newblock {\em Phys. Lett. B}, 213:471, 1988.

\bibitem{imbo}
T.~D. Imbo, C.~S. Imbo, and C.~S. Sudarshan.
\newblock Identical particles, exotic statistics and braid groups.
\newblock {\em Phys. Lett. B}, 234:103, 1990.

\bibitem{pra}
J.~Jacak.
\newblock Application of the path integral quantization to indistinguishable
  particle systems topologically confined by a magnetic field.
\newblock {\em Phys. Rev. A}, 97:012108, 2018.

\bibitem{nat}
P.~{\L}yd{\.z}ba, L.~Jacak, and J.~Jacak.
\newblock Hierarchy of fillings for the {FQHE} in monolayer graphene.
\newblock {\em Sci. Rep.}, 5:14287, 2015.

\bibitem{nature}
J.~Jacak.
\newblock Unconventional fractional quantum {H}all effect in bilayer graphene.
\newblock {\em Sci. Rep.}, 7:8720, 2017.

\bibitem{jet}
J.~Jacak and L.~Jacak.
\newblock The commensurability condition and fractional quantum hall effect
  hierarchy in higher {L}andau levels.
\newblock {\em JETP Letters}, 102:19–--25, 2015.

\bibitem{montecarlo1}
Orion Ciftja and Carlos Wexler.
\newblock Monte carlo simulation method for laughlin-like states in a disk
  geometry.
\newblock {\em Phys. Rev. B}, 67:075304, 2003.

\bibitem{laughlin2}
R.~B. Laughlin.
\newblock Anomalous quantum {H}all effect: an incompressible quantum fluid with
  fractionally charged excitations.
\newblock {\em Phys. Rev. Lett.}, 50:1395, 1983.

\bibitem{jain}
J.~K. Jain.
\newblock Composite-fermion approach for the fractional quantum {H}all effect.
\newblock {\em Phys. Rev. Lett.}, 63:199, 1989.

\bibitem{sr1a}
J.~Jacak.
\newblock Unconventional fractional quantum {H}all effect in bilayer graphene.
\newblock {\em Sci. Rep. Supplementary Information}, 7:8720(1--14), 2017.

\bibitem{bil}
D.~K. Ki, V.~I. Falko, D.~A. Abanin, and A.~Morpurgo.
\newblock Observation of even denominator fractional quantum {H}all effect in
  suspended bilayer graphene.
\newblock {\em Nano Lett.}, 14:2135, 2014.

\bibitem{bil1}
Georgi Diankov, Chi-Te Liang, FranĂ§ois Amet, Patrick Gallagher, Menyoung Lee,
  Andrew~J. Bestwick, Kevin Tharratt, William Coniglio, Jan Jaroszynski, Kenji
  Watanabe, Takashi Taniguchi, and David Goldhaber-Gordon.
\newblock Robust fractional quantum hall effect in the n=2 landau level in
  bilayer graphene.
\newblock {\em Nature Comm.}, 7:13908, 2016.

\bibitem{sk1}
Y.~W. Suen, L.~W. Engel, M.~B. Santos, M.~Shayegan, and D.~C. Tsui.
\newblock Observation of a {$\nu =1/2$} fractional quantum {H}all state in a
  double-layer electron system.
\newblock {\em Phys. Rev. Lett.}, 68:1379, 1992.

\bibitem{sk2}
P.~Eisenstein, G.~S. Boebinger, L.~N. Pfeiffer, K.~W. West, and Song He.
\newblock New fractional quantum {H}all state in double-layer two-dimensional
  electron systems.
\newblock {\em Phys. Rev. Lett.}, 68:1383, 1992.

\bibitem{amet}
F.~Amet, A.~J. Bestwick, J.~R. Williams, L.~Balicas, K.~Watanabe, T.~Taniguchi,
  and D.~{Goldhaber-Gordon}.
\newblock Composite fermions and broken symmetries in graphene.
\newblock {\em Nat. Commun.}, 6(6838), 2015.

\bibitem{pajac}
T.~Kramer.
\newblock Two interacting electrons in a magnetic field: comparison of
  semiclassical, quantum, and variational solutions.
\newblock {\em AIP Conf. Proc.}, 1323:178, 2010.
\newblock arXiv:1009.6051.

\bibitem{pajac1}
M.~Taut.
\newblock Two electrons in a homogeneous magnetic field: particular analytical
  solutions.
\newblock {\em Journal of Physics A}, 27:1045, 1994.

\bibitem{pajac3}
U.~Merkt, J.~Huser, and M.~Wagner.
\newblock Energy spectra of two electrons in a harmonic quantum dot.
\newblock {\em Phys. Rev. B}, 43:7320, 1991.

\bibitem{pajac2}
A.~Mathew and M.~K. Nandy.
\newblock Two electrons in a cylindrical quantum dot under constant magnetic
  field.
\newblock {\em Physica B}, 421:127, 2013.

\bibitem{pc}
P.~{\L}yd{\.z}ba and J.~Jacak.
\newblock Identifying particle correlations in quantum {H}all regime.
\newblock {\em Annalen der Physik}, 49:1700221, 2017.

\bibitem{epl}
J.~Jacak and L.~Jacak.
\newblock Recovery of {L}aughlin correlations with cyclotron braids.
\newblock {\em EPL}, 92:60002, 2010.

\bibitem{pan2003}
W.~Pan, H.~L. St{\"o}rmer, D.~C. Tsui, L.~N. Pfeiffer, K.~W. Baldwin, and K.~W.
  West.
\newblock Fractional quantum {H}all effect of composite fermions.
\newblock {\em Phys. Rev. Lett.}, 90:016801, 2003.

\bibitem{jain2007}
J.~K. Jain.
\newblock {\em Composite Fermions}.
\newblock Cambridge UP, Cambridge, 2007.

\bibitem{ll}
L.~Jacak J.~Jacak.
\newblock Commensurability condition and fractional quantum {H}all effect
  hierarchy in higher {L}andau levels.
\newblock {\em J. Exp. Theor. Phys. Letters}, 102:19, 2015.

\bibitem{halp}
B.~I. Halperin.
\newblock Theory of the quantized {H}all conductance.
\newblock {\em Helv. Phys. Acta}, 56:75, 1983.

\bibitem{montecarlo2}
R.~Morf and B.~I. Halperin.
\newblock Monte carlo evaluation of trial wavefunctions for the fractional
  quantized hall effect: Spherical geometry.
\newblock {\em Z. Phys. B Condensed Matter}, 68:391, 1987.

\bibitem{metropolis}
N.~Metropolis, A.~W. Rosenbluth, M.~N. Rosenbluth, A.~M. Teller, and E.~Teller.
\newblock Equation of state calculations by fast computing machines.
\newblock {\em J. Chem. Phys.}, 1953.

\end{thebibliography}
\end{document}